# Reexamining Machine Learning Models on Predicting Thermoelectric Properties

Chung T. Ma[1,*], and S. Joseph Poon[1,2]

1 Department of Physics, University of Virginia, Charlottesville, VA 22904, USA
2 Department of Material Science and Engineering, University of Virginia, Charlottesville, VA 22904, USA

## Abstract

Thermoelectric (TE) materials can generate clean energy by transforming waste heat into electricity. The effectiveness of thermoelectric materials is measured by the dimensionless figure of merit, ZT. The quest for high-ZT materials has drawn extensive research experimentally and theoretically. However, due to the vast material space, finding high-ZT materials is time-consuming and costly. To improve the efficiency of discovering new thermoelectric materials, recent studies have employed machine learning (ML) with databases to search for high-ZT candidates. In this work, we examine the effects of adding various physical concepts on the performance of machine learning models in predicting TE properties. The objective is to improve the model's ability to capture the underlying physics in designing TE materials. These concepts include short-range order and crystal structure class. Results show some improvements in accuracy. However, the current models do not distinguish between dilute alloys and concentrated alloys, rendering them inadequate in predicting doping effects. To better capture the electronic band structure effect from doping, we included various dopants' properties as features. This increases the prediction accuracy in doped materials. Furthermore, we used a genetic algorithm to rank features for various thermoelectric properties to provide physical insight into key parameters in designing thermoelectric materials.

## Introduction

In energy generation, a significant portion of the input energy is lost in heat [1]. Converting this waste heat into electricity can greatly improve the overall efficiency of producing energy globally. Thermoelectric (TE) technology is a promising approach that makes use of this waste heat and turns it into electricity [2-5]. The performance of thermoelectric materials is determined by the dimensionless figure of merit, ZT, which is defined as $ZT = (S^2\sigma/\kappa)T$, where S is the Seebeck coefficient, $\sigma$ is the conductivity, $\kappa$ is the sum of electronic $\kappa_e$ and lattice thermal conductivity $\kappa_l$, and T is the temperature [1-5]. A wide range of materials have been investigated for thermoelectric applications, including Half-Heusler compounds [6-9], $Bi_2Te_3$-based alloys [10-12], GeTe-based alloys [13-15], PbTe-based alloys [16-18], chalcogenides [19-21], oxide [22-24], and compounds based on Mg with Si or Sb [25-27]. With such a large variety of candidate materials to choose from, the design and the discovery of high-ZT materials has been a prolonged and expansive process. While both experimental and theoretical works have advanced the development of thermoelectric materials, high-throughput and cost-effective methods are needed to accelerate breakthroughs.

To this end, recent studies have turned to machine learning (ML) as a cost-effective way to search for high-ZT materials in this vast material space. In these studies, databases of thermoelectric materials have been constructed using both experimental data and first-principles calculations [28-40]. A broad array of raw features are used in modeling, such as elemental properties, atomic structure, and grain size. Feature engineering techniques [29,30], including auto-encoder and random forest [30], have been applied to process these raw features to enhance the model performance. Different kinds of regression models include Decision Tree Regression (DTR) [29-31,34,35], Gaussian process regression (GPR) [32-33], least absolute shrinkage and selection operator (LASSO) algorithm [34-36], neural network (NN) [30,31,34,35,37-39], support vector regression (SVR) [35,40], and gradient boosting machine (GBM) [29-31,36], have been developed. These models have been used to predict relevant thermoelectric properties, such as band gap, carrier concentration, electrical conductivity, Seebeck coefficient, power factor (PF), lattice thermal conductivity, and figure of merit ZT. Some of these models have demonstrated excellent predictive ability, achieving $R^2$ of over 0.90, and identifying candidates of high-ZT materials. Furthermore, classification models have predicted the formation of various intermetallic phases, including Heusler, B2, and BCC phases [41-47]. Some of these models have achieved accuracy of over 90% in phase predictions.

In this study, we focus on predicting various thermoelectric properties and reexamine the machine learning models previously developed [29-35] by incorporating additional alloy features as well as improving the interpretability of the model. The original model employed an auto-encoder to compress 146 raw features, followed by using GBM to train a regression model [48]. To better capture underlying physics, we also incorporated crystal structure and short-range order in the ML models. While these improve the model accuracy in predicting ZT, further work is needed in predicting doped materials with dopant concentrations below 5 atomic percentages. Above 5 atomic percentages, the dopant is considered a "principal element" in the alloy [49]. We developed a doping model, loosely based on adopting 5 atomic percent as the crossover from doping to alloying to better capture the impact of doping on electronic band structures. In this model, dopant elemental properties are included as features and the model shows noticeable improvement in predicting ZT of doped materials. Moreover, we applied the genetic algorithm to rank feature importance concerning ZT. This improved the interpretability of the machine learning models. Thus, besides enhancing the accuracy of state-of-the-art machine learning models in predicting TE properties, the present investigation also provides valuable insight in designing thermoelectric materials.

**Method**

In this study, we utilized a dataset of 14100 thermoelectric alloys with the corresponding ZTs measured at various temperatures. Most of the raw features are extracted from the Magpie database in Matminer [50]. These features include minimum, maximum, range, weighted compositional average, and deviation of elemental properties such as atomic number, atomic weight, Electronegativity, and band-gap energy of T = 0 K ground state. We refer to this original model as the "alloy solution model" herein, as this model treats each thermoelectric alloy like a solution of these elemental properties. For short-range order, the mixing enthalpy $H_{mix}$ is calculated using Miedema's model as follows [51].

$$H_{mix} = \left( \Sigma_{(i \neq j)} c_i c_j e^{(-\Delta H_{(i,j)}/k_B T)} \right) / \Sigma_{(i \neq j)} c_i c_j \quad (1)$$

Where $c_i$ and $c_j$ are the atomic concentrations of elements i and j, respectively, and $H_{(i,j)}$ is the mixing enthalpy of the i-j elemental pair obtained from Miedema's model.

For doped materials, we extended the raw features to include 36 dopant-specific elemental properties.

We consider this model the "specialized doping model". For preprocessing, a min-max algorithm is applied to convert features into arrays of vectors ranging from 0 to 1. A five-layer auto-encoder neural network model with a rectified linear unit (ReLU) activation function using Adam optimizer [52] is applied to encode the raw features. Adam Optimizer uses combination of root mean square propagation and stochastic gradient descent with momentum for optimization, in order to code the each layer of features/ Once coded, the LightGBM [47] algorithm is employed to construct decision trees and make predictions for ZT, Seebeck coefficient S and thermal conductivity κ. During the regression training, all data and coded features carry equation weight and decision trees are built sequentially using residual, which is the difference from actual and predicted value, from previous decision tree through the LightGBM algorithm. Figure 1 shows a schematic of the workflow of this machine learning model. Additionally, a genetic algorithm, coupled with random forest, is implemented to rank the importance of the raw features relative ZT.

**Results and Discussions**

First, we evaluated the alloy solution model with the original 146 raw features. Table 1 shows the $R^2$ value of this model in predicting ZT. The model achieves $R^2$ of 0.86. While this shows the model performs reasonably well, further examinations revealed some shortcomings. As shown in Figure 2, the predicted ZT of $Ge_{1-x}Sb_xTe$ (with x = 0.08) by the alloy solution model shows great agreement with the experimental data [15] at temperatures up to 500 K. However, for temperatures above 500 K, the predicted ZT deviates from the measured ZT. At 600 K, the predicted ZT is approximately 20% lower than experimental value. The discrepancy persists at 700 K and 750 K. We also looked into a different class of thermoelectric materials, as shown in Figure 2. Comparing predicted ZT and measured ZT [25] of $Mg_{3.2-y}Mn_ySb_{1.5}Bi_{0.49}Se_{0.01}$ (with y = 0.05) reveals a deviation of 20% to 30% the temperature range from 300 K to 650 K. For these thermoelectric materials, while the model can capture the overall trend of temperature-dependent ZT, discrepancies remain. To address these issues, we introduced physical concepts, such as short-range order and crystal structure class to improve the model's ability to capture the underlying physics.

We incorporated short-range order into the alloy solution model through mixing enthalpy $H_{mix}$ from eq. (1) as additional raw features. From Table 1, the result shows a small increase of $R^2$, from 0.86 to 0.87, compared to the original alloy solution model. Improvement in ZT prediction is seen for both $Ge_{1-x}Sb_xTe$ (with x = 0.08) and $Mg_{3.2-y}Mn_ySb_{1.5}Bi_{0.49}Se_{0.01}$ (with y = 0.05). Figure 2 shows that, similar to the original alloy solution model, the alloy solution model with short-range order can accurately predict ZT of $Ge_{1-x}Sb_xTe$ (with x = 0.08) up to 500 K. However, at 500 K and above, the model with short-range order data shows marginal improvements, compared to the original alloy solution model. At 600 K, the deviation decreases from approximately 20% to about 15%. The improvement is more pronounced at 750 K. The difference is down to nearly 10%, half of that of the original model. For $Mg_{3.2-y}Mn_ySb_{1.5}Bi_{0.49}Se_{0.01}$ (with y = 0.05), the model with short-range order shows consistent improvement in predicting ZT across the temperature range from 300 K to 650 K. The deviation from experimental ZT decreases to 15%-20%, compared to 20%-30% by the original model. Overall, introducing short-range order in the alloy solution model improves ZT prediction accuracy.

On the other hand, we implemented crystal structure class by separating the dataset grouos of thermoelectric materials accordingly. As shown in Table 1, this modification results in an increase in $R^2$ from 0.86 to 0.875 compared to the original alloy solution model. In predicting ZT of $Ge_{1-x}Sb_xTe$ (with x = 0.08), Figure 2 shows the same performance by the alloy solution model with crystal structure and the original model for temperatures up to 500 K. However, at higher temperatures (>500 K), the model model with crystal structure is more accurate. The discrepancy decreases to less than 5% at 750 K, compared to about 20% for the original model. For predicting the ZT of $Mg_{3.2-y}Mn_ySb_{1.5}Bi_{0.49}Se_{0.01}$ (with y = 0.05), Figure 3 shows improvement in ZT predictions by the model with crystal structure throughout the temperature range of 300 K to 650 K. The deviation in predicted ZT decreases to around 10% to 15%, compared to 20%-30% by the original model. Since improvements are found in both the inclusion of short-range order and crystal structure, we combined short-range order in the alloy solution model with the crystal structure. The $R^2$ value of this model in predicting ZT increases to 0.88. Moreover, the predicted ZT move closer to the measured ZT for the alloys shown in Figure 2 and Figure 3 Thus, incorporating physical concepts such as short-range order and crystal structure improves the model accuracy in predicting ZT.

To provide a more thorough understanding of designing thermoelectric materials, we examined the ability of the alloy solution in predicting other thermoelectric properties. We used the alloy solution model to predict the Seebeck coefficient S and thermal conductivity κ, which are more closely connected to the band structure of a given alloy than ZT. Figure 4 and 5 shows the comparison of experimental and predicted S and κ in $Ge_{1-x}Sb_xTe$ (with x = 0.08) and $Mg_{3.2-y}Mn_ySb_{1.5}Bi_{0.49}Se_{0.01}$ (with y = 0.05), respectively. In Figure 4 (a), predictions of S in $Ge_{1-x}Sb_xTe$ (with x = 0.08) by alloy solution model show deviations from experiments for temperatures from 300 K to 750 K. At 300 K, the difference between predictions and experiments is near 30%, and this difference remains at ~20% to ~30% throughout these temperatures. In order to improve the model's accuracy we implemented short-range order and crystal structure in the alloy solution model. First, with only addition of short-range order, a marginal improvement in S predictions is observed for temperatures from 300 K to 750 K. For example, at 300 K, the difference between predictions and experiments decreases to about 25%, comparing to ~30% by the original model. Very similar trend is found by separating the dataset into different classes of crystal structures, where only small improvements in S predictions are observed. Further improvements in S predictions are observed by introducing both short-range order and crystal structure into the original alloy solution model. At 300 K, the predicted S is about 15% smaller than the measured S. At 750 K, the predicted S is about 10% smaller. Throughout 300 K to 750 K, using the alloy solution with short-range order and crystal structure, the discrepancies in S predictions decreases to ~10% to ~15%, from ~20% to ~30%.

On the other hand, Figure 5 (a) shows the predictions of S in $Mg_{3.2-y}Mn_ySb_{1.5}Bi_{0.49}Se_{0.01}$ (with y = 0.05). Using the original alloy solution model, the predicted |S| is ~10% to ~20% smaller than the measured |S| between 300 K to 600 K. For example, at 300 K, the predicted |S| is ~10% smaller, while at 600 K, the predicted |S| is ~20% smaller, compare to experiment. With short-range order or crystal structure, the alloy solution model improves marginally. The largest improvement is found with the alloy solution model with short-range order and crystal structure, which is the same trend as those observed in ZT predictions and |S| predictions of $Ge_{1-x}Sb_xTe$ (with x = 0.08). Here, the discrepancies in |S| predictions reduces to ~5% to ~10%, compare to ~10% to ~20% originally. Looking at the entire dataset, as shown in Table 1, the $R^2$ value of the original model in predicting |S| is 0.8. With the implementation of short-range order or crystal structure in the original model, the $R^2$ value in predicting |S| increases to 0.815 and 0.82, respectively. This matches the marginal improvement seen in Figure 4 and Figure 5. The $R^2$ value in predicting |S| further improves to 0.825 in the alloy solution model with short-range order and crystal structure. Therefore, by incorporating short-range order and crystal structure improves the model's ability in predicting |S|.

Next, we examined the predictions of $\kappa$ by these models. Figure 4 (b) and 5 (b) shows the comparison of experimental and predicted $\kappa$ in $Ge_{1-x}Sb_xTe$ (with x = 0.08) and $Mg_{3.2-y}Mn_ySb_{1.5}Bi_{0.49}Se_{0.01}$ (with y = 0.05), respectively. For $Ge_{1-x}Sb_xTe$ (with x = 0.08), the original model over-predicted $\kappa$ for temperatures from 300 K to 750 K. At 300 K, the predicted $\kappa$ is nearly 30% larger than experiment's. The difference reduces to ~5% at a higher temperature for 750 K. For alloy solution model with short-range order or crystal structure, very little improvement is observed. The predicted $\kappa$ from these models almost overlap with the original model. Similarly, for alloy solution model with short-range order and crystal structure, the improvement is still rather tiny. At 300 K, the predicted $\kappa$ is nearly 25% larger than measured $\kappa$, compare to ~30% by the original model. The difference reduces to ~5% at a higher temperature for 750 K, which is comparable to the original model. For $Mg_{3.2-y}Mn_ySb_{1.5}Bi_{0.49}Se_{0.01}$ (with y = 0.05), the effect of implementing short-range order and crystal structure in the model is more pronounced. In the original alloy solution model, the predicted $\kappa$ is about 15% larger than measured $\kappa$ at 400 K. This difference remains less than 15% for higher temperatures up to 675 K. In particular, the discrepancy reduced to almost 0% at 600 K and 675 K. For both alloy solution model with short-range order and alloy solution model with crystal structure, the predicted $\kappa$ is about 5% larger than measured $\kappa$ at 400 K, an improvement from ~15% in the original model. At 600 K and 675 K, both model's predicted $\kappa$ pretty much overlap with predicted $\kappa$ by the original model, which is very close to measured $\kappa$. Further improvements are seen in predicted $\kappa$ by alloy solution model with short-range order and crystal structure. The discrepancies is less than 3% throughout 300 K to 675 K. For the entire dataset, as shown in Table 1, the $R^2$ value in predicting $\kappa$ improves marginally by implementing short-range order and crystal structure. Using the original alloy solution model, the $R^2$ value in predicting $\kappa$ is 0.84. The $R^2$ value in predicting $\kappa$ improves to 0.85 in both alloy solution model with short-range order and alloy solution model with crystal structure. The $R^2$ value in predicting $\kappa$ further improves to 0.855 in both alloy solution model with both short-range order and crystal structure. This trend is similar to those seen in predicting ZT and S. Therefore, these confirm that by introducing these physical concepts in this machine learning model, the model can better capture the underlying physics in designing thermoelectric materials and make better predictions in various thermoelectric properties.

| Properties | Alloy solution model | Alloy solution model with short range order | Alloy solution model with crystal structure | Alloy solution model with short range order and crystal structure |
|---|---|---|---|---|
| ZT | 0.86 | 0.87 | 0.875 | 0.88 |
| S | 0.80 | 0.815 | 0.82 | 0.825 |
| $\kappa$ | 0.84 | 0.85 | 0.85 | 0.855 |

Table 1 Comparison of $R^2$ values in ZT, S and $\kappa$ predictions by alloy solution model, alloy solution model with short-range order, alloy solution model with crystal structure, and alloy solution model with short-range order and crystal structure.

A deeper examination reveals the model's failure in predicting ZT for certain doped materials, particularly for alloys with small dopant atomic percentages. Some predicted ZT deviates from experiments greatly. An example of such deviations is shown in Figure 6 (a). Experiment data of $Hf_{0.3}Zr_{0.7}Co(Sn_{0.3}Sb_{0.7})_{1-x}Al_x$ shows that the ZT increases significantly, by a factor of two, in alloys with Al concentration (x) between 0.01 to 0.02 due to resonant doping [9]. However, the alloy solution model fails to capture this dramatic increase in ZT. While the model predicts ZT reasonably well for the undoped alloy, the predicted ZT of doped alloys is over 50% lower than the measured ZT. For example, at x = 0.015 and 900 K, the measured ZT is just above 1.4, whereas the model predicts a value close to 0.8, which is over 40% lower. This difference persists for all doping level of $Hf_{0.3}Zr_{0.7}Co(Sn_{0.3}Sb_{0.7})_{1-x}Al_x$, and for temperatures ranging from 300 K to 900 K. When considered only doped materials, the $R^2$ values for the alloy solution model drops to 0.6, as shown in Table 2. Therefore, a specialized model in doping is needed to enhance the model's ability to accurately predict ZT for doped materials.

| Properties | Alloy solution model | Specialized doping model | Specialized doping model with dataset of ZT increase by >20% |
|---|---|---|---|
| ZT | 0.60 | 0.70 | 0.73 |
| S | 0.57 | 0.63 | 0.65 |
| κ | 0.68 | 0.73 | 0.75 |

**Table 2** Comparison of alloy solution model and specialized doping model $R^2$ value in predicting ZT, S, and κ of doped materials.

Similarly, large discrepancies is found in S and κ predictions for doped materials. Figure 7 (b) and (c) shows the measured and predicted S and κ in $Hf_{0.3}Zr_{0.7}Co(Sn_{0.3}Sb_{0.7})_{1-x}Al_x$, respectively. For undoped alloy, the S prediction matches experiments well for temperatures ranging from 300 K to 500 K. For temperatures above 500 K, discrepancies increases from ~5% at 550 K to ~10 % at 900 K. The differences between predicted and measured S are far greater in Al-doped $Hf_{0.3}Zr_{0.7}Co(Sn_{0.3}Sb_{0.7})_{1-x}Al_x$. For x = 0.005, while predicted S agrees with measured S at 300 K and 400 K, disagreements are seen at 500 K and above, going from ~5% difference at 500 K to ~20 % difference at 900 K. Moreover, the models preform far worse for x = 0.01, 0.015 and 0.02. At 300 K, the discrepancy between predicted S and measured S is ~25% and increase to over 30% at 900 K. This large disagreement for x = 0.01, 0.015 and 0.02 is comparable to those found in ZT predictions. Similar behavior is also seen in predictions of κ. The predicted κ matches experiments for undoped $Hf_{0.3}Zr_{0.7}Co(Sn_{0.3}Sb_{0.7})_{1-x}Al_x$ throughout 300 K to 900 K. For all Al-doped alloys (x = 0.005, 0.01, 0.015, and 0.02), discrepancies are observed for all temperatures. In particular, the difference between predicted κ and measured κ is almost 40% for x = 0.005 and 0.01 at 300 K. Thus, these doped materials present a barrier in predicting thermoelectric properties for the alloy solution model.

To overcome the challenge in predicting thermoelectric properties of doped alloys, we developed a specialized doping model. In this model, the elemental features of the dopants are incorporated as raw features to better capture the electronic band structure effects by doping. Table 2 summarizes the improvement in ZT prediction by employing the specialized doping model. The $R^2$ increases from 0.6 in the original model to 0.7. An example of improvement is shown in Figure 7, which presents the same experimental alloy $Hf_{0.3}Zr_{0.7}Co(Sn_{0.3}Sb_{0.7})_{1-x}Al_x$ as in Figure 3. Using the specialized doping model, the predicted ZT values are closer to the experimental data. The discrepancy reduces to about 30%, compared to nearly 50% by the original model. For instance, at x = 0.015 and 900 K, the predicted ZT is about 1.1, which is nearly 30% lower than the measured ZT of ~1.4. In contrast, the original alloy solution model predicted a ZT nearly 40% lower than the measured ZT. This improvement in predicted ZT is seen in all Al doping concentrations (x = 0.005 to 0.02) and across temperatures from 300 K to 900 K.

Similar improvements is observed in S and κ predictions for doped materials. Table 2 summarizes the improvement in S and κ prediction using the specialized doping model. In S predictions, the alloy solution model achieves $R^2$ value of 0.57. $R^2$ value in S predictions improves to 0.63 using the specialized doping model. For predicting κ, the alloy solution model achieves $R^2$ value of 0.68, while the specialized doping model achieves $R^2$ value of 0.73. We examined this more carefully using $Hf_{0.3}Zr_{0.7}Co(Sn_{0.3}Sb_{0.7})_{1-x}Al_x$ as an example. Figure 8 (b) and (c) shows the measured and predicted S and κ by the specialized doping model in $Hf_{0.3}Zr_{0.7}Co(Sn_{0.3}Sb_{0.7})_{1-x}Al_x$, respectively. For x = 0.005, the specialized doping model shows good agreement in predicting S with experiments throughout temperature range from 300 K to 900 K, where alloy solution model only shows agreements for temperatures up to 500 K. For x = 0.01, 0.015, and 0.02, the improvements in predicting S are more significant. At 300 K, the predicted S by the specialized doping model is ~10% smaller than measured S, compare to ~25% smaller using the alloy solution model. At 900 K, using the specialized doping model, the differences between S predictions and measurements are ~20%, compare to almost 30% using alloy solution model. A comparable improvement is seen for predicting κ using the specialized doping model. For example, at 300 K, the predicted κ by the specialized doping model is ~20% larger than measured κ, compare to almost 40% larger using alloy solution model. These results show that even though some disagreements remain between predicted thermoelectric properties in doped materials and experiments, the specialized doping perform better overall than the alloy solution model in predicting ZT, S and κ in doped materials.

To further examine the persisting deviations in predicting ZT of doped materials, histograms of ZT distribution for doped materials are plotted in Figure 9. These histograms reveal discrepancies in both low and high ZT ranges. For example, the alloy solution model predicts nearly 30% of doped materials with ZT between 0 to 0.1, while experiments find only about 20% of doped materials fall in the this range. Although the specialized doping model is more accurate, it still predicts about 25% doped materials with ZT between 0 to 0.1, which is above the experimental value of 20%. Furthermore, both the alloy solution model and specialized doping model predict less high ZT materials (ZT > 1.4), compared to experimental observations. These results show that discrepancies remain in predicting ZT for doped materials, so further improvements in this specialized doping are needed to enhance accuracy.

To address the underestimation of ZT by both models we introduced a modified specialized doping model. In this model, we filtered the dataset by including only doped materials that show a 20% or more increase in ZT compared to the respective undoped alloys. By including only doped alloys with a significant ZT enhancement, we target to improve the model's ability to capture the effects of doping on electronic band structure. As shown in Figure 8, which presents the same experimental alloy $Hf_{0.3}Zr_{0.7}Co(Sn_{0.3}Sb_{0.7})_{1-x}Al_x$ as in Figure 6 and Figure 7, the predicted ZT by the new model is closer to the experimental ZT. For example, at x = 0.015 and 900 K, the predicted ZT is approximately 1.3, which is only 10% lower than the measured ZT of ~1.4. This shows a significant improvement compared to both the alloy solution model (Figure 6) and the original specialized doping model (Figure 7). Furthermore, the improvements in ZT predictions of this alloy are consistent throughout the temperature range of 300 K to 900 K. Upon inspection of the $R^2$ value in Table 2, the modified model shows an $R^2$ value of 0.73 for predicting ZT, increases from 0.6 in the original alloys solution model and 0.7 in the original doping model. For predicting S, the modified model shows an $R^2$ value of 0.65 for predicting S, increases from 0.57 in the original alloys solution model and 0.63 in the original doping model. Similarly, improvement is seen in κ predictions. The modified model shows an $R^2$ value of 0.65 for predicting κ, increases from 0.57 in the original alloy solution model and 0.63 in the original doping model. This means that by filtering dataset in the doping model, the model can predict ZT, S and κ better than the original alloy solution model.

Moreover, as shown in Figure 8, the modified model shows better agreement with experimental data in ZT distributions compared to the other models. It predicts fewer materials low-ZT and more high-ZT materials (ZT > 1.4), which is more in line with measured results. However, some discrepancies remain between predicted and experimental ZT. For example, the modified model predicts about 15% of materials with ZT between 0 and 0.1, but experiments show around 20%. These discrepancies could be related to the phase formation effects in doped materials, where impurities or phase separations lead to a dramatic change in ZT [10,16,18]. Such phase effects are not captured by the current model, so further development is needed to improve prediction accuracy of doped materials.

| Figure-of-Merit ZT | Seebeck Coefficient S | Thermal Conductivity κ |
|---|---|---|
| Temperature | Temperature | Temperature |
| Electronegativity (maximum) | Mendeleev number (standard deviation) | Valence d-electron (mean) |
| Melting T (mean) | Column (mean) | space group (mean) |
| Electronegativity (minimum) | Electronegativity (min) | Melting T (Max) |
| Covalent radius (mean) | Covalent radius (mean) | Valence d-electron (standard deviation) |
| Ground state volume per atom (mean) | Mendeleev Number (mean) | Electronegativity (mean) |
| Mendeleev Number (mean) | Ground state volume per atom (mean) | Atomic Weight (standard deviation) |
| Electronegativity (mean) | Atomic number (mean) | Melting T (mean) |
| Mendeleev number (deviation) | Melting T (mean) | Mendeleev number (standard deviation) |
| Melting T (mode) | Ground state volume per atom (standard deviation) | Valence f-electron (mean) |

Table 3 Comparison of top 10 most important features with respect to Figure-of-Merit ZT, Seebeck Coefficient S and thermal conductivity κ, obtained from genetic algorithm.

Finally, to interpret the machine learning model, we implemented a genetic algorithm to rank the each raw features with respect to ZT. Table 3 lists the top 10 most important features. The highest-ranking feature is temperature, which is trivial since ZT, by definition, is proportional to temperature T. Other key elemental features include Electronegativity, melting temperature, covalent radius, ground state atomic volume, and Mendeleev number. These features play certain roles in crafting the electronic structure of the alloys. They affect the atomic packing, overlap of atomic orbitals, electron density, overlap of electron clouds, and charge transfer, which affect hybridization and bandgaps in semiconductors. The Mendeleev number, which holds the trends in chemical properties such as ionization energy, further substantiate the effect of these features. For S, the top features mostly overlaps with those for ZT, with the exception of atomic number and column. This makes sense physically as both ZT and S has dependence on the band structures of the alloys. For κ, there are more notable differences. While temperature remain top feature, the number of valance f and d electrons and space group of the elements appear in the top 10 features for κ, compared to those for ZT and S. This is likely due to the electrons and lattice structures play a more prominent role in thermal conductivity, which by definition, is the sum of electronic and lattice contributions. These findings provide a deeper understanding of the models, which could further accelerate designing novel thermoelectric materials. On the hand, the bottom 10 features, which are the least important, are all related to the number of unfilled or valance s-electrons. This is because, for all elements used in thermoelectric alloys in this dataset, the maximum valance s-electrons are 2 or the minimum unfilled electrons are 0. More importantly, from an alloy designing point of view, the outer shells electrons, specifically the p-, d-, and f- electrons, have a far greater influence on material properties than the s-electrons. Therefore, it makes physical sense that the raw elemental features related to the s-electrons are the least important with respect to ZT.

**Conclusion**

We have enhanced the existing machine learning model for predicting key thermoelectric properties by incorporating key physics concept. The incorporation of short-range order and crystal structure show some improvements in the accuracy in figure of merit ZT, Seebeck coefficient S, and thermal conductivity κ predictions. In addition, we have further developed the model tailored for doped materials. This model noticeably outperforms the original alloy solution model in predicting ZT, S and κ. We also employed a genetic algorithm to rank features with respect to ZT. By connecting the machine learning model with the underlying physics of thermoelectric materials, this work paves the way for designing high-ZT materials in future thermoelectric applications.

**List of figures**

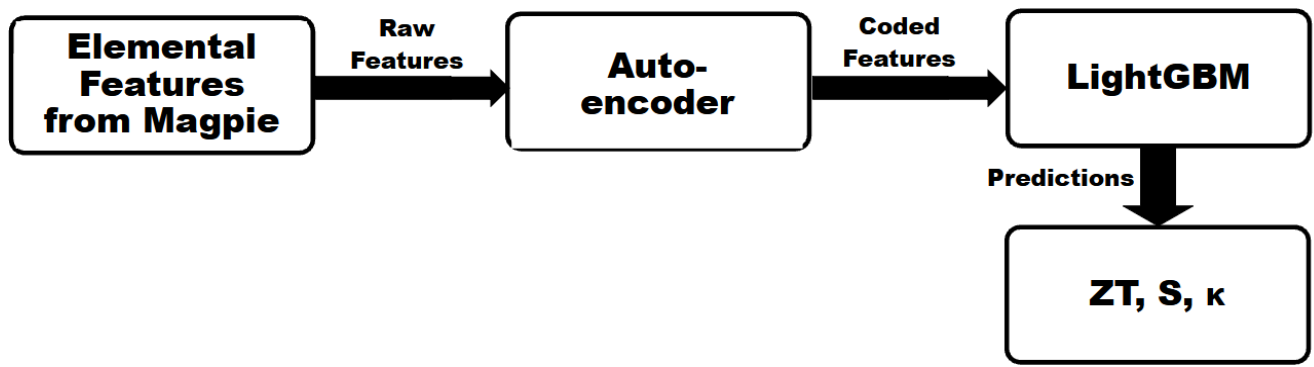

**Figure 1** Schematic of the machine learning workflow employed in this study.

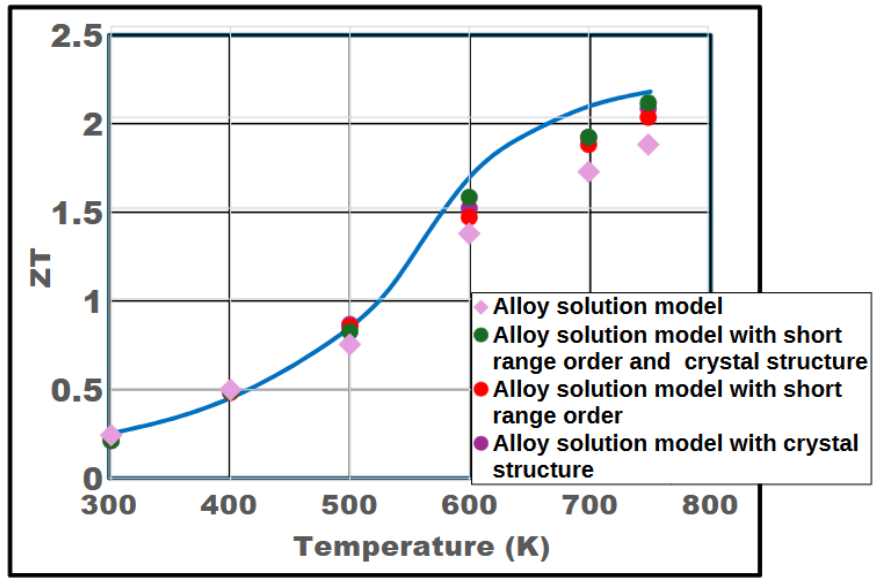

**Figure 2** Comparison of measured ZT (blue line) [15] and predicted ZT by alloy solution model (pink), alloy solution model with short-range order (red), alloy solution model with crystal structure (purple), and alloy solution model with short-range order and crystal structure (dark green) of $Ge_{1-x}Sb_xTe$ with $x = 0.08$.

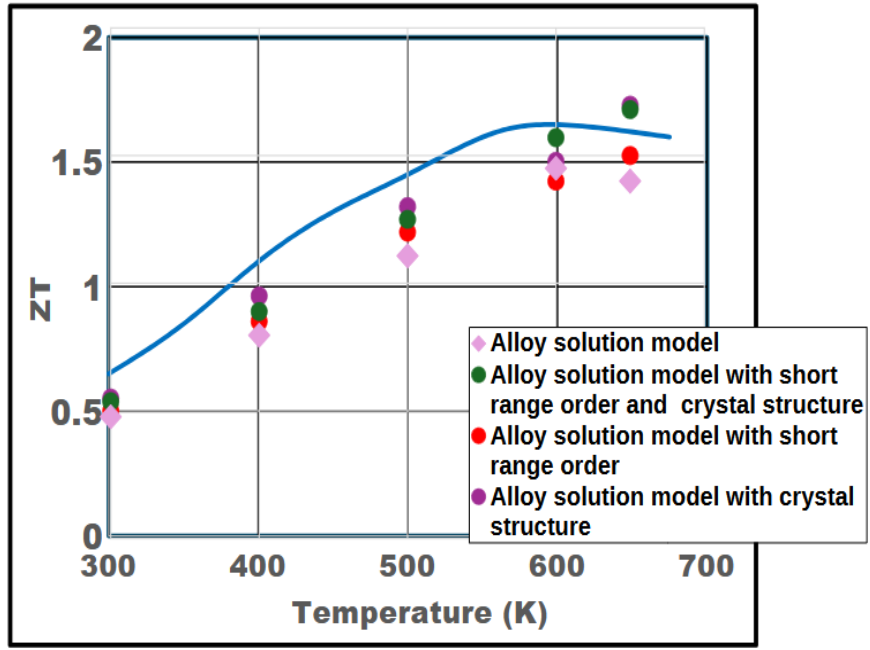

**Figure 3** Comparison of measured ZT (blue line) [25] and predicted ZT by alloy solution model (pink), alloy solution model with short-range order (red), alloy solution model with crystal structure (purple), and alloy solution model with short-range order and crystal structure (dark green) of $Mg_{3.2-y}Mn_ySb_{1.5}Bi_{0.49}Se_{0.01}$ with y = 0.05.

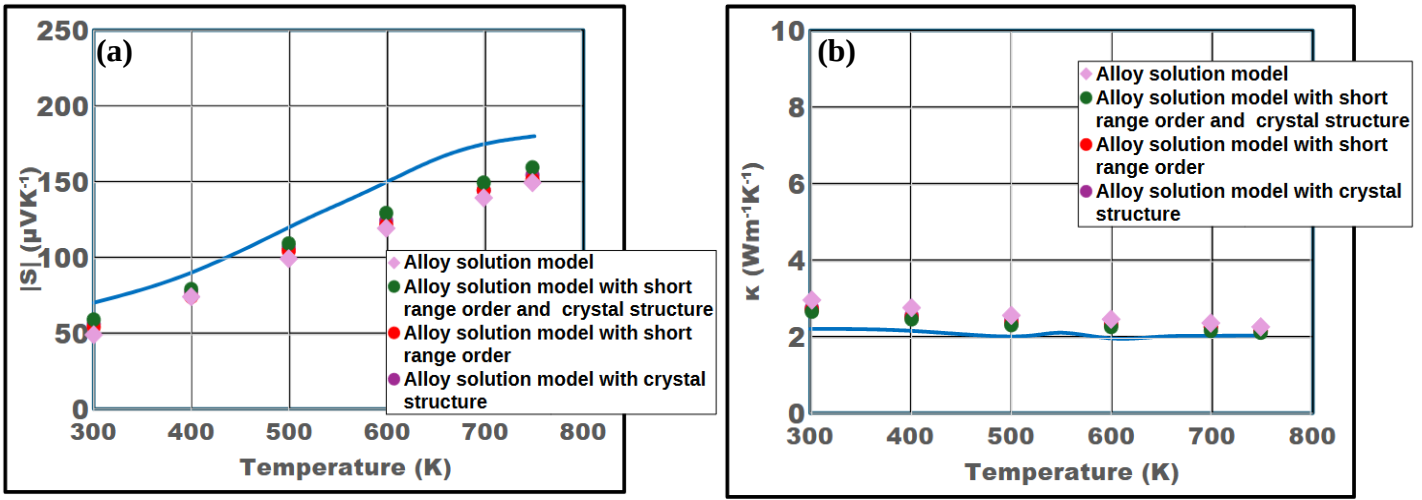

**Figure 4 (a)** Comparison of measured S (blue line) [15] and predicted S by alloy solution model (pink), alloy solution model with short-range order (red), alloy solution model with crystal structure (purple), and alloy solution model with short-range order and crystal structure (dark green) of $Ge_{1-x}Sb_xTe$ with $x = 0.08$. **(b)** Comparison of measured κ (blue line) [15] and predicted κ by alloy solution model (pink), alloy solution model with short-range order (red), alloy solution model with crystal structure (purple), and alloy solution model with short-range order and crystal structure (dark green) of $Ge_{1-x}Sb_xTe$ with $x = 0.08$.

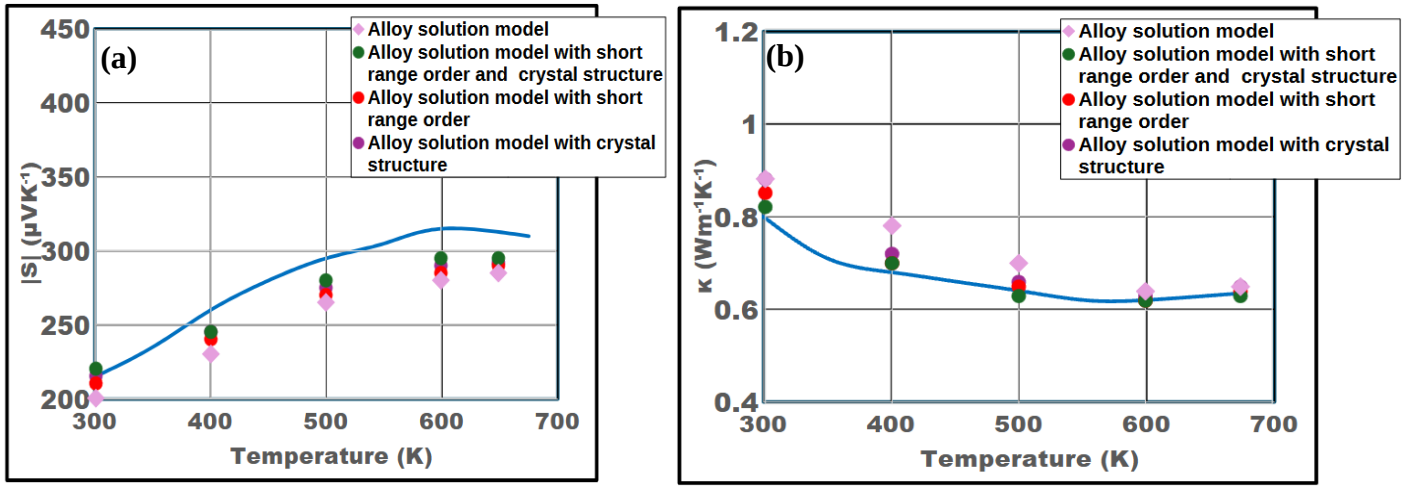

**Figure 5 (a)** Comparison of measured S (blue line) {25} and predicted S by alloy solution model (pink), alloy solution model with short-range order (red), alloy solution model with crystal structure (purple), and alloy solution model with short-range order and crystal structure (dark green) of $Mg_{3.2-y}Mn_ySb_{1.5}Bi_{0.49}Se_{0.01}$ with y = 0.05. **(b)** Comparison of measured κ (blue line) [25] and predicted κ by alloy solution model (pink), alloy solution model with short-range order (red), alloy solution model with crystal structure (purple), and alloy solution model with short-range order and crystal structure (dark green) of $Mg_{3.2-y}Mn_ySb_{1.5}Bi_{0.49}Se_{0.01}$ with y = 0.05.

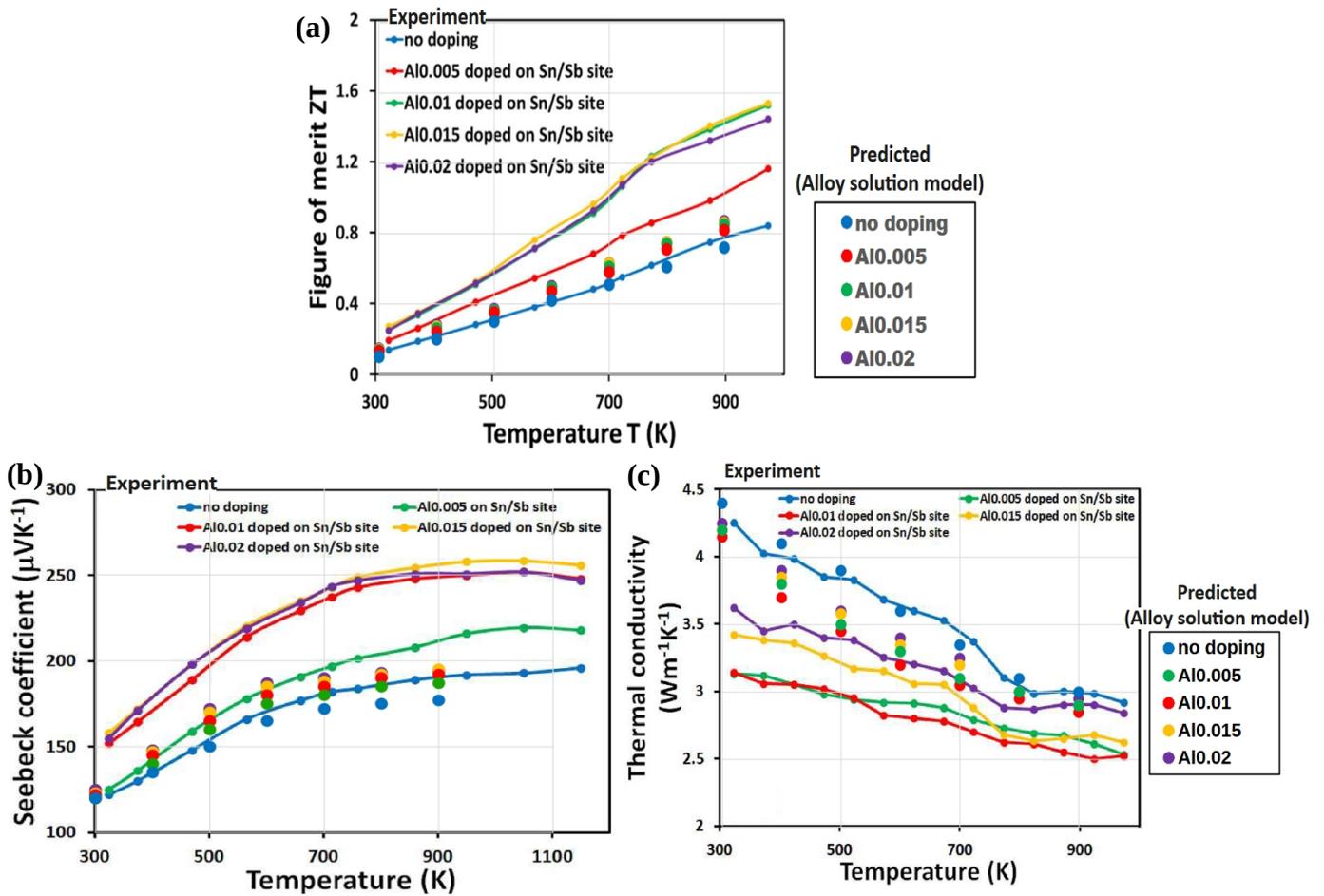

**Figure 6 (a)** Comparison of measured ZT [9] and predicted ZT by alloy solution model of a doped material $Hf_{0.3}Zr_{0.7}Co\,(Sn_{0.3}Sb_{0.7})_{1-x}Al_x$ with various Al concentration. **(b)** Comparison of measured S [9] and predicted S. **(c)** Comparison of measured κ [9] and predicted κ.

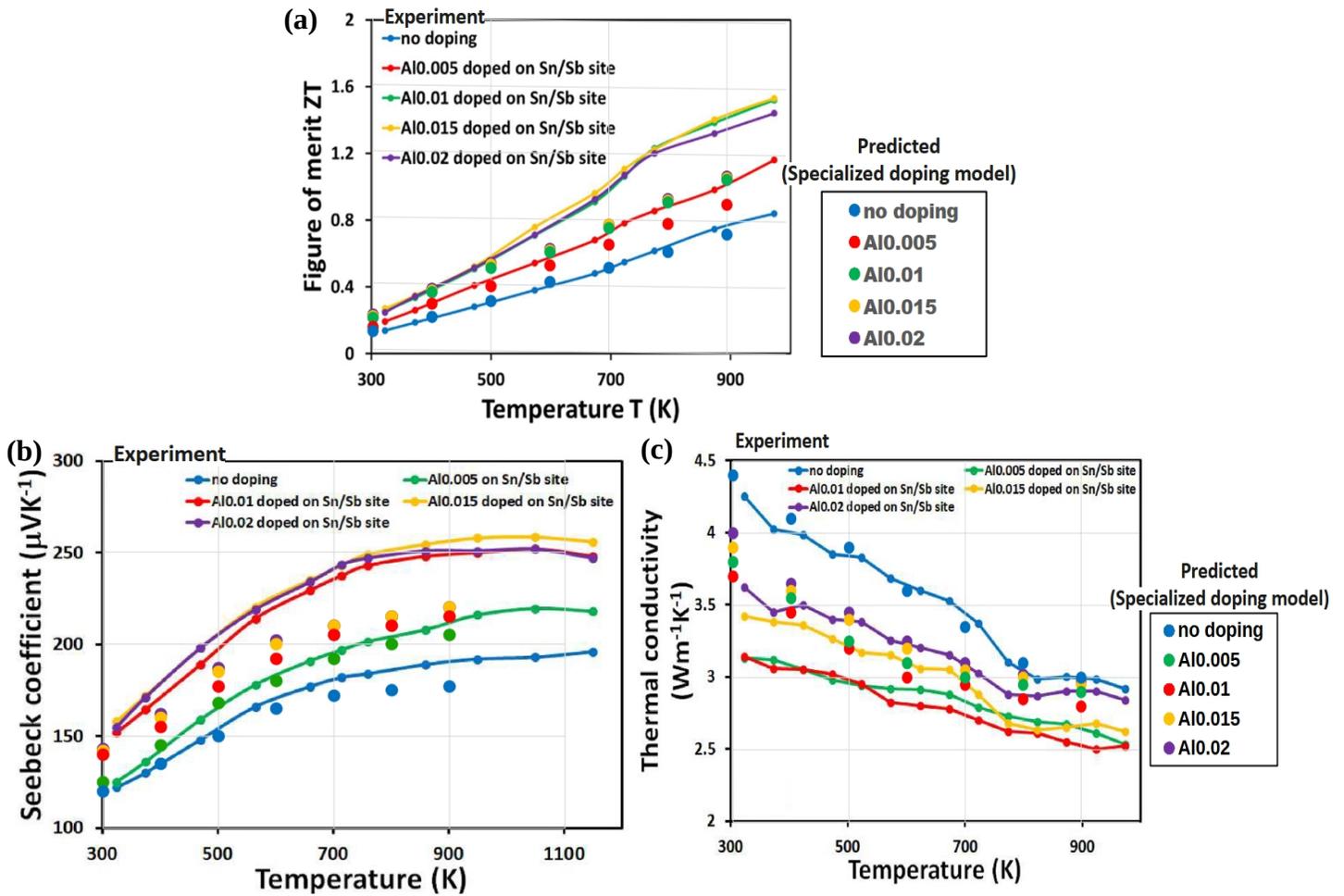

**Figure 7 (a)** Comparison of measured ZT [9] and predicted ZT by specialized doping model of a doped material $Hf_{0.3}Zr_{0.7}Co\,(Sn_{0.3}Sb_{0.7})_{1-x}Al_x$ with various Al concentration x. **(b)** Comparison of measured S [9] and predicted S. **(c)** Comparison of measured κ [9] and predicted κ.

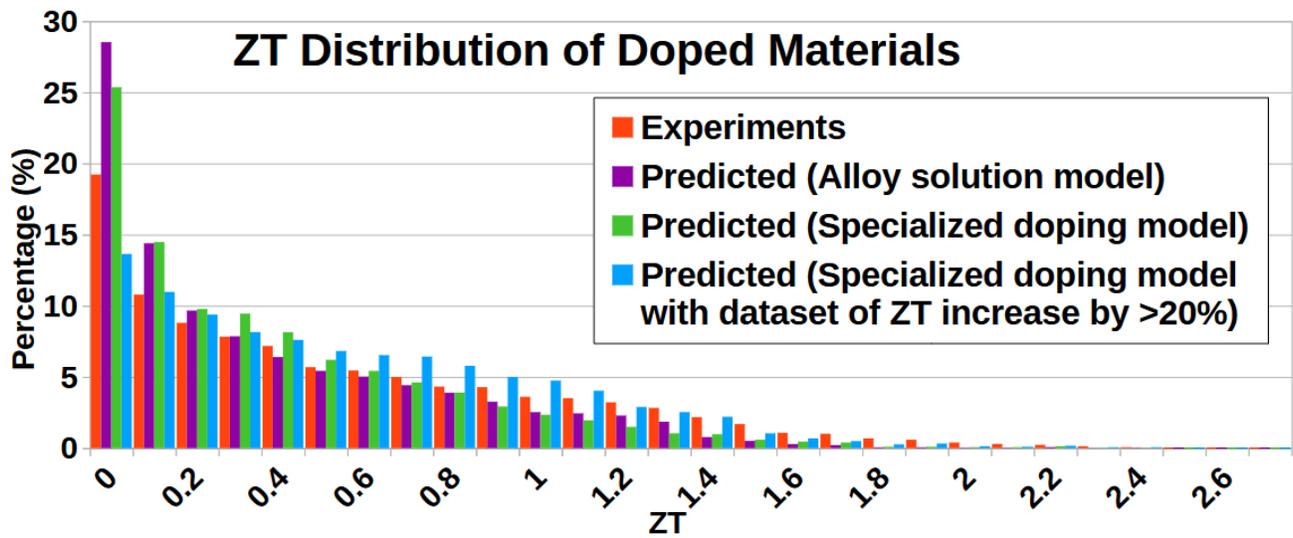

**Figure 8** ZT distribution of doped materials from experiments (red), prediction by alloy solution model (purple), prediction by specialized doping model (green)., and prediction by specialized doping model with dataset of ZT increase by greater than 20% (blue).

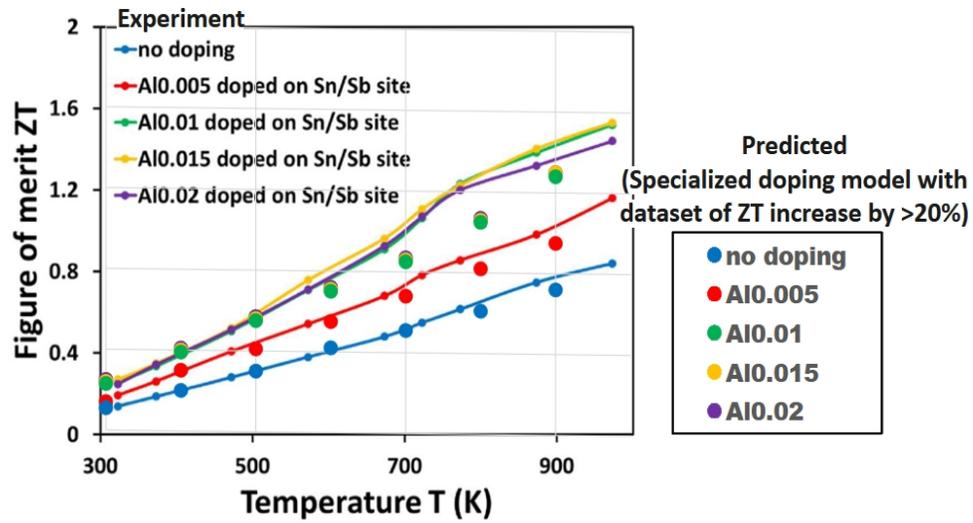

**Figure 9** Comparison of measured ZT [9] and predicted ZT by specialized doping model with dataset of ZT increase by greater than 20% of a doped material $Hf_{0.3}Zr_{0.7}Co\,(Sn_{0.3}Sb_{0.7})_{1-x}Al_x$ with various Al concentration x.